\documentclass[twocolumn,showpacs,preprintnumbers,amsmath,amssymb,prl]{revtex4}

\usepackage{graphicx}
\usepackage{dcolumn}
\usepackage{bm}

\begin{document}

\preprint{Phys. Rev. Lett. {\bf 102}, 037003 (2009)}

\title{
Giant Coupling of Fe-spin and the As-As Hybridization  in Iron-Pnictides}


\author{T. Yildirim$^{1,2}$}\email{taner@nist.gov}%
\affiliation{%
$^{1}$NIST Center for Neutron Research, National Institute of Standards and
Technology, Gaithersburg, Maryland 20899, USA
\\$^{2}$Department of Materials Science and Engineering, University of
Pennsylvania, Philadelphia, PA 19104, USA}%

\date{Phys. Rev. Lett. {\bf 102}, 037003 (2009)}

\begin{abstract} 
From first principles calculations we unravel surprisingly strong
interactions between arsenic ions in iron-pnictides, the strength of which  is  
controlled by the Fe-spin state. 
Reducing the Fe-magnetic moment, weakens the Fe-As bonding, and in turn, 
increases As-As interactions, causing giant reduction in the c-axis.
For CaFe$_2$As$_2$ system, this reduction is as large as 1.4 \AA.
Since the large c-reduction 
has been recently observed only under high-pressure\cite{cTphase}, 
our results suggest that
the iron magnetic moment should be present in Fe-pnictides at all times
at ambient pressure. Finally, the conventional electron-phonon coupling in the
collapsed phase of CaFe$_2$As$_2$ gives a maximum  $T_c$ of 0.6 K 
and can not explain the $\sim12$~K superconductivity observed in some experiments. 
Implications of these findings on 
the mechanism of  superconductivity in iron-pnictides are discussed.

\end{abstract}

\maketitle

The recent discovery of superconductivity at 
T$_c$'s up to 55~K in iron-pnictide systems\cite{kamihara,sm_43k,ce_41k,pr_52k}
has sparked enormous interest in this class of materials. So far two types of materials
have been discovered. The first one is the rare-earth  pnictide oxide layered systems, 
REOFeAs which is denoted as "1111"\cite{kamihara,sm_43k,ce_41k,pr_52k,afe2as2_refeas}. 
The second class is the so called "122"
systems with the chemical formula  
MFe$_2$As$_2$ (M=Ca,Sr, etc)\cite{bafe2as2,srfe2as2,cafe2as2,srfe2as2_sc,afe2as2_refeas}. 
The 122 systems are
 simpler in terms of their crystal structure due to absence of REO-layers.
  The crystal structure of CaFe$_2$As$_2$ 
is shown in the inset to Fig.~1.

 
 The pressure-induced superconductivity in these system  is 
 particularly interesting\cite{pres_cafe2as2_sc,sr_ba_fe2as2_pres,pres_cafe2as2}
 because it   provides a new avenue for investigation of the mechanism of the high-T$_c$ 
 superconductivity.
 Very recently a high-pressure neutron scattering study has reported that
 the CaFe$_2$As$_2$ system undergoes a surprising transition to 
 a "collapsed tetragonal phase"
 (cT-phase) under applied pressure in which the c-parameter is reduced 
 from 11.7 \AA~ to 10.6 \AA\cite{cTphase}. 
 Motivated by this interesting report, in this letter we present
  a detailed  first-principles study of Fe-pnictides 
 with many surprising results.


\begin{figure}
\includegraphics[width=7.5cm]{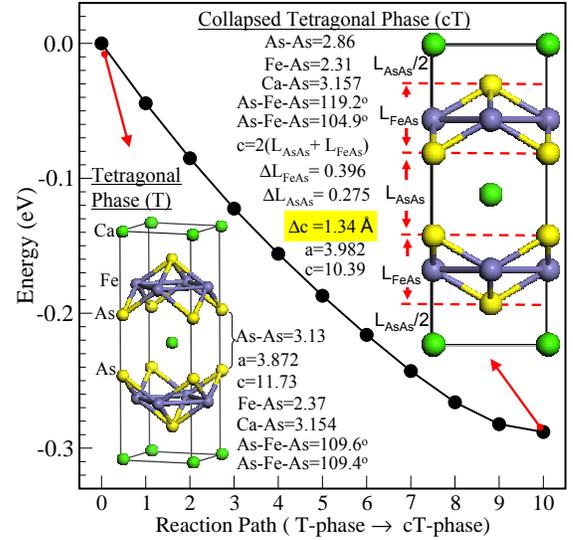} 
\caption{
(color online)
Total energy as the Ca122 tetragonal phase goes  to
collapsed-tetragonal phase without any energy barrier during non-spin polarized
structural optimization.  The insets show the initial (T-Phase) and final (cT-phase) 
Ca122 structures with relevant bond-distances (in \AA) and angles (in degrees). 
}
\label{fig1}
\end{figure}

 \begin{table}
\caption{Various optimized  structural parameters
for NM, AF2, and AF1 spin configurations, respectively. 
The experimental data are taken 
from Refs[\onlinecite{cafe2as2,canaxfe2as2,bafeas,cruz}].
$^*$The AF1 configuration goes to NM for Ca$_{0.5}$Na$_{0.5}$Fe$_2$As$_2$ .}
\begin{center}
\begin{tabular}{|c|c|c|c|c|c|c||c|} \hline \hline
   & a & b & c & As(z) &  d$_{FeAs}$ &  M$_{Fe}$ &E (meV) \\ \hline
  \multicolumn{8}{ |c|}{CaFe$2$As$_2$}\\  \hline
NM & 5.63 & 5.63  & {\bf 10.39} &  0.36251  & 2.309  & 0& 0.0\\
AF1 & 5.65  & 5.65 & {\bf 10.60} & 0.36440  & 2.338  & 1.3 & -16\\
AF2 & 5.61 & 5.48  & 11.61 &       0.36695  & 2.367  & 2.2  & -100\\
Exp. & 5.68  & 5.68 & {\bf 11.73} &0.3665 & 2.370  & 1.0 & --\\ \hline
 \multicolumn{8}{| c|}{Ca$_{0.5}$Na$_{0.5}$Fe$_2$As$_2$}\\ \hline 
NM & 5.59  & 5.59 & 10.52 &  0.36284 &   2.31  & 0  & 0.0\\
AF1$^*$ & 5.59 & 5.59  & 10.52 &  0.36284 &  2.31  & 0 & 0.0\\
AF2 & 5.43  & 5.53 & 12.05 &  0.36536  & 2.382  & 2.4  & -97  \\
Exp. & 5.42 & 5.42  & 11.86 &  -- &  --      & 0.0   &  -- \\ \hline
  \multicolumn{8}{ |c|}{BaFe$2$As$_2$}\\  \hline
NM & 5.58 & 5.58  &  12.45 &  0.3479 &    2.319    & 0 & 0.0\\
AF1 & 5.64  & 5.64 & 12.73 &  0.35231 &   2.382    & 2.1 &   -80 \\
AF2 & 5.70 & 5.59  & 12.83 &  0.3549    & 2.408  & 2.4 & -169 \\
Exp.& 5.52 & 5.52 &  13.02 &  0.3545  &   2.397  & 1.0 & -- \\ \hline
  \multicolumn{8}{ |c|}{LaOFeAs}\\  \hline
NM & 5.64 & 5.64  & 8.59  &   0.35944     &  2.332  & 0       & 0.0 \\
AF1 & 5

.69  & 5.69  & 8.71 &  0.35128  &     2.393  & 2.1   & -86\\
AF2 & 5.67  &  5.73 & 8.72&   0.34860    &   2.409  & 2.4   & -190 \\
Exp. & 5.70  & 5.70 & 8.737&  0.3479  &      2.407  & 0.35 &  --\\ \hline
\end{tabular}
\end{center}
\end{table}

  We discovered that the optimized  c-lattice
 parameter of CaFe$_2$As$_2$  varies by about 1.5 \AA~ depending on the
 magnetic configuration considered. Such a giant coupling of spin-state of an ion
 with its lattice is almost unheard of and deserves  detailed investigation.
Reducing Fe-moment by about 1/2 gives lattice
 parameters  that are quite close to
  high-pressure data\cite{cTphase}.
  Hence, the collapsed-phase  does not necessarily mean the total
  loss of Fe magnetism as suggested
  in ref.[\onlinecite{cTphase}] but  could be due to large
  reduction of the Fe-moment. 
 We explain this strange behavior by noting that in iron-pnictide systems 
 there are significant 
 As-As hybridization (both intra- and inter-plane arsenic ions) 
 whose strength is controlled by the
 Fe-spin state. Reducing the Fe-spin state reduces the Fe-As bonding,which in turn, increases the 
 As-As bonding and causes the observed huge reduction in the lattice parameters.
  This effect is 
 maximized in the case of Ca122 system due to close proximity of two arsenic ions in adjacent Fe-planes
 (see Fig. 1). We argue that since there
is no observation of large c-axis reduction during the normal to superconducting 
phase transition in iron-pnictides, the iron magnetic moment should be present  at all times
at ambient pressure. Otherwise one should have seen a large reduction in the c-axis as  
recently reported from high-pressure neutron scattering experiments\cite{cTphase}. 
Hence it seems that the Fe-magnetic moment in its paramagnetic state
 (i.e. no long range order but
non-zero Fe-site moment) is either required for the superconductivity or at least is not
detrimental to  the superconductivity.

The first-principles calculations were performed within the plane-wave
implementation of the Perdew-Burke-Ernzerhof  (PBE) 
generalized gradient approximation (GGA) 
 to density functional theory  as implemented in the PWSCF package\cite{pwscf}. 
The details of the parameters used in the calculations can be found in
Refs.\onlinecite{yildirim_prl1,ca122_condmat}.
In order to get a general understanding of the iron-pnictides, in this study we consider one example
of each class of pnictides; namely CaFe$_2$As$_2$ for the 122 system with the smallest Ca-ion available and
the BaFe2As2 with the largest metal Ba. For the 1111 system, we consider LaOFeAs. 
We also study a doped
122-system, i.e. Na$_{0.5}$Ca$_{0.5}$Fe$_2$As$_2$. Since in our 
$\sqrt{2}\times\sqrt{2}$-cell we have
four chemical formula, we consider a supercell where two Na and two Ca are ordered. For each given
system, we have performed full structural optimization  including the lattice
parameters and the atomic positions. We consider our optimization is converged when the maximum force on
each atom is less than 0.005 eV/\AA~ and the pressure is less than 0.1 kbar.
We have performed the full structural optimization for non-magnetic (NM), i.e. "non-spin polarized",
checkerboard antiferromagnetic (AF1) and stripe (AF2) spin configurations. Our results are summarized
in Table~1.  As expected, the ground state for all four systems is the stripe AF2 phase and the
optimized parameters are in good agreement with the experimental data at ambient conditions.

The most striking and surprising finding listed in Table~1 is the giant dependence of the
optimized c-lattice parameter on the spin-configuration considered. For the case of CaFe$_2$As$_2$,
we note that AF1 configuration is the next stable  state (after the AF2) but the c-value is
significantly reduced; 11.63 \AA~ versus 10.60 \AA~ for AF2 and AF1 spin configurations, respectively.
The difference is even larger, when the Fe-magnetism is ignored (i.e. non-spin polarized calculations).
The optimized c-value for NM-state is 10.39 \AA, which is 1.34 \AA~  shorter than the experimental value
at ambient pressure. 
We note that the optimized  lattice parameters, a=5.65 \AA~ and c=10.6 \AA~ for the 
AF1 phase are in reasonable agreement with the neutron data (a=5.8 \AA~ and c=10.6 \AA)\cite{cTphase}. 
Hence, based on this result, it is premature to conclude that the observed experimental 
cT-phase is  due to the complete disappearance of the Fe-magnetism but 
as we shall see below 
it is closely related to 
the magnitude of the Fe-spin.

\begin{figure}
\includegraphics[width=6.0cm]{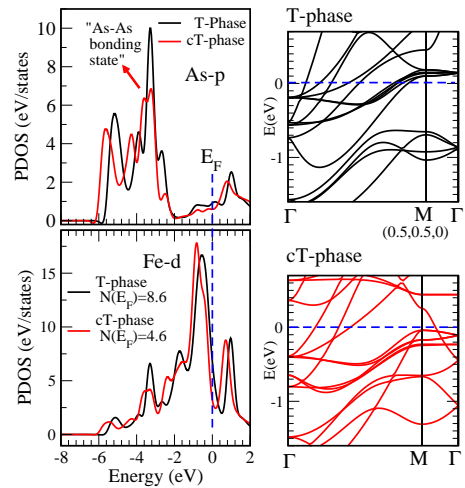}
\caption{
(color online)
The projected electronic density of states (PDOS) for As p-orbital (top-left) and
Fe-d orbital (bottom left) for T- (black) and cT-phases (red). 
The right two panels show the band structure along 
$\Gamma$-M-$\Gamma$.
The As-As $p_z$-bonding orbital is indicated by the arrow on the top left panel
and it is shown in Fig.~3 }
\label{fig2}
\end{figure}

In Fig.~1 we show that   the T-phase directly collapses into
the cT-phase without any energy barrier if the Fe-magnetic moment  is ignored.
During the c-axis collapse, there is significant and comparable 
decrease in the height of the Fe-As
and As-Ca-As planes, indicating that the whole lattice
almost uniformly shrinks. 
Since there is no energy barrier between the
T-phase and the cT-phase, as soon as we  loose the Fe-magnetic moment for
some reason, we should see the expected c-lattice reduction. Since this does not seem to happen in the
superconducting samples, it is tempting to conclude that we have the Fe-moment present in the
superconducting phase. This is a quite important result by itself to understand the mechanism of the
high T$_c$ superconductivity observed in iron pnictides.


In order to get a better understanding of the cT-phase, we have performed electronic band
structure calculations and studied the projected atomic density of states.
Our results for the T-phase agrees well with previous calculations\cite{singh,singh2,mazin_review,veronica}.  
Fig.~2 shows that despite to
the 1.4 \AA~  c-reduction, the band structure and the PDOS of both phases are surprisingly similar. 
 The main difference between the
two phases is that most of the bands are shifted in lower energy in the cT-phase which is expected
since the cT-phase has the lower energy. This downward shift of the bands is most obvious in the band
structure plot along the $\Gamma-M$ direction as shown in Fig.~2. In the T-phase, there were several
states above the Fermi level which crosses the E$_F$ along the $M-\Gamma$ direction. This results in a larger
density of states at the Fermi level. However for the case of cT-phase, we note that most of the bands
just above  E$_F$ in the T-phase are now just below the E$_F$ in the cT-phas and there is only one
band which crosses the E$_F$ along the $\Gamma-M$ direction. This explains why  N$(E_F)$ is reduced
in the cT-phase.  The other and probably the most important difference is the change in the peak shape
of the states near 3 eV below  E$_F$. Visual inspection of these orbitals indicate that there is
significant As-As hybridization in this system. The As-As hybridization becomes more significant in the
cT-phase. This observation is quite unexpected and  and as we shall see below it actually explains
the mystery why the Fe-spin state has a huge effect on the As-z coordinates  as first noted in 
[\onlinecite{picket}].

\begin{figure}
\includegraphics[width=8cm]{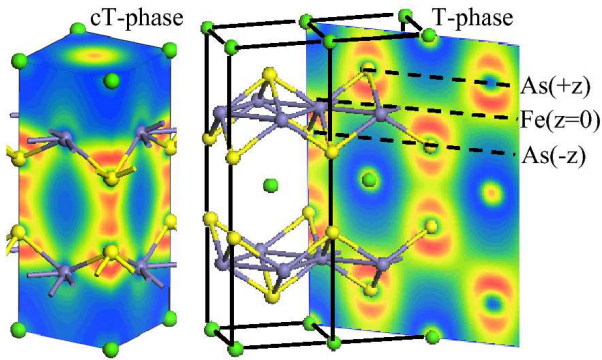}
\caption{
(color online)
Contour plot of the  As-As $p_z$-hybridization orbital for
cT-phase  (left) and T-phase  (right), respectively. 
Note that the As-As hybridization present in both phases is 
much more significant in the cT-phase. There 
is also clear hybridization between intra-As atoms 
below and above the Fe-plane in the T-phase (right).   
 }
\label{fig3}
\end{figure}

In order to demonstrate that there are large hybridization between As ions in the Ca122 system, we show
the contour plots of the relevant orbitals in Fig.~3. 
It is clear that the As ion below the top Fe-plane makes a bond (or
hybridizes) with the arsenic ion which is above the lower Fe-plane. Hence this overlap of the As-As
along the c-axis makes this system quite isotropic and far from being layered system. 
According to bond-population
analysis, the As-As bond  strength  increased almost twice in the cT-phase. 
 Due to close proximity of the As ions in adjacent
Fe-layers, the observation of the As-As interaction is probably not that surprising. What is surprising is  
to see that there is almost the same type of hybridization between two arsenic ions on the same Fe-plane
as shown in the right panel of Fig.3.

Since we have shown that the As ion
above the Fe-plane has a strong overlap with the As ion below the same iron plane, their interaction is
automatically increased as the Fe-As interaction decreases due to decrease in the Fe-moment which
changes the chemistry of the Fe ion. Therefore, we have now a mechanism which explains why the As ion z-values
get shorter with the decreasing Fe-moment. Our mechanism also explains why we see a smaller reduction 
in the c-axis for the LaOFeAs  than the 122 system as listed in Table~1. 
The reduction in the c-axis  in the LaOFeAs system is due to the intra-plane As-As
interaction only since there are no two adjacent FeAs planes to interact to with each
other as in the case of Ca122. Our theory also predicts that for larger ions like Ba, we should see less
c-reduction because the As-As distance between two adjacent planes are now larger due to larger ionic
radius of Ba. In Table~1, we also show that similar c-reduction with Fe-spin occurs in the doped
Na$_{0.5}$Ca$_{0.5}$Fe$_2$As$_2$ system  as well.

\begin{figure}
\includegraphics[width=6.0cm]{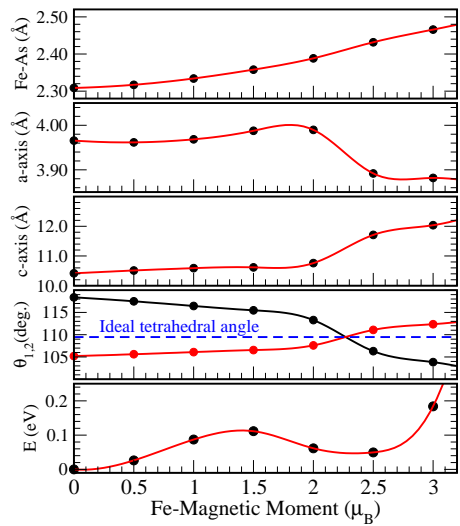}
\caption{
(color online)
Various structural parameters as a function of Fe-moment from fixed-spin
 calculations with ferromagnetic
iron spin configuration.
 }
\label{fig4}
\end{figure}

In order to convince the reader further that the spin-state of the iron is the key parameter that
controls the As-As bonding in these systems, we have performed fixed spin calculations  
for CaFe$_2$As$_2$ and
the results are shown in Fig.~4. 
We consider ferro spin configuration to show
that the spin-structure is a secondary effect and the main effect is the on-site Fe-spin state. 
As we see from Fig.~4, when the Fe-moment is zero, the c-axis is the smallest consistent with the
strongest As-As interaction (because As-Fe bonding is weak for non-magnetic Fe). 
As the Fe moment becomes significant, the Fe-As interaction gets stronger and
therefore the As-As distance starts to increase to optimize the Fe-As bonding. Ironically at the
Fe-moment of $\mu_B$=2.2, we have the ideal tetragonal arrangement of the four As ion around each
iron during which the two As-Fe-As angles are equal. At this point, the low- to high-Fe spin transition
is obvious during which the a-axis is reduced while  c-axis is increased significantly. 

Since our results suggest that Fe-magnetism is either totally lost or reduced by half 
in the cT-phase, one wonders if the observed $\sim12$~K superconductivity in the 
vicinity of the collapse cT-phase of  CaFe$_2$As$_2$\cite{pres_cafe2as2}
can be explained by conventional electron-phonon (e-ph) coupling? 
In order to address this question, we have calculated 
phonon spectrum and Eliashberg function from linear response theory\cite{pwscf}. 
We used basically the same method and the equivalent parameters 
that are used in Ref.\onlinecite{boeri}  for LaOFeAs. 
Our results are summarized in Fig.~5 and very 
similar to those for LaOFeAs. We obtained a value of 
electron-phonon coupling $\lambda = 0.23$ and the  logarithmically average frequency 
$\omega_{log}= 218$~K, which gives $T_c=0.6$~K using the Allen-Dynes formula with
$\mu^{*}=0$ (i.e. an upper bound for T$_c$).
Hence, the mechanism of superconductivity 
in the cT-phase of CaFe$_2$As$_2$ is likely unconventional and probably
it is the same as in the other iron-pnictides. 
This finding together with our results summarized in Table~1 suggest 
that  in the cT-phase the Fe-magnetism is not totally lost  with pressure and but 
it is partially reduced.  Hence it seems that there is an optimum strength of 
Fe-spin state  that is required for high-T$_c$ superconductivity.

\begin{figure}
\includegraphics[width=6.5cm]{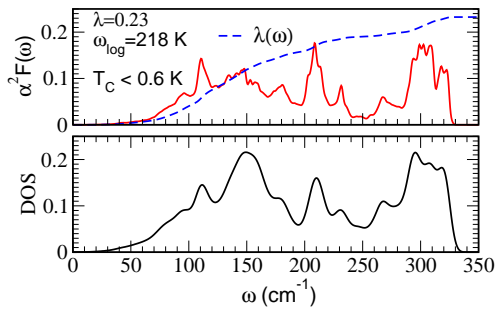}
\caption{
(color online)
Phonon density of states (DOS), Eliashberg function ($\alpha^2F(\omega)$) and the
frequency-dependent e-ph coupling $\lambda(\omega)$ (dashed line) for CaFe$_2$As$_2$ 
in the cT-phase. 
 }
\label{fig5}
\end{figure}

In conclusion, we have revealed surprisingly strong As-As interactions in iron-pnictides.
The strength of this interaction is 
controlled by the Fe-As chemical bonding. 
Reducing the Fe-moment, reduces the Fe-As bonding, 
which in turn increases the As-As interaction along
the z-axis, causing arsenic atoms  on opposite sides of 
Fe-square lattice to move towards   each other. 
This  explains the
high sensitivity of the z-atom positions and the large reduction of the c-axis with 
the loss of Fe-magnetic moment.
We showed that  under external pressure, the high Fe-spin AF2 structure (i.e. stripe phase)
should transform to a new structure  with  low Fe-spin state  and
 significantly reduced c-axis.  We think that this could
be the phase recently observed by high-pressure neutron scattering\cite{cTphase}. 
This is also 
consistent with our finding that the pressured induced $\sim12$~K superconductivity can not
be explained by the conventional e-ph coupling. The Fe-magnetism is still needed for an
unconventional mechanism.
Since at ambient pressure, we do
not observe large c-axis drops in the superconducting samples, we
conclude that the Fe-magnetic moment
should be present at all times in these systems, 
at least in 122  materials such as CaFe$_2$As$_2$. 
The  giant coupling of the on-site 
Fe-magnetic moment with the As-As bonding that we have discovered here may provide 
a mechanism for the superconductivity.  

The author acknowledges useful discussions with M. A. Green.

\end{document}